\documentclass[prl,twocolumn,showpacs]{revtex4}
\usepackage{amsthm}
\usepackage{amsmath}
\usepackage{amssymb}
\usepackage{amsfonts}
\usepackage{epsf}
\usepackage{upref}

\IfFileExists{myowntimes.sty}{\usepackage[rmboldsymbol]{myowntimes}}{
\usepackage{times}
\usepackage{mathrsfs}
}

\newtheoremstyle{thm}{2ex}{2ex}{\itshape\rmfamily}{}{\bfseries\rmfamily}
{}{1.7ex}{}
\newtheoremstyle{rem}{1.3ex}{1.3ex}{\rmfamily}{}{\itshape\rmfamily}
{}{1.5ex}{}

\theoremstyle{thm}

\newtheorem*{Main Theorem}{Main Theorem.}

\theoremstyle{rem}

\newcommand{\eff}{\operatorname{eff}}

\begin{document}

\title{
General Theory of Lee-Yang Zeros in Models with First-Order Phase
Transitions}

\author{M.~Biskup$^*$, C.~Borgs$^*$, J.T.~Chayes$^*$,
L.J.~Kleinwaks$^\dag$, R.~Koteck\'y$^\ddag$}
\affiliation{$^*$Microsoft Research, One Microsoft Way, Redmond WA 98052, U.S.A.}
\affiliation{$^\dag$Department of Physics, Princeton University, Princeton NJ
08544, U.S.A.}
\affiliation{$^\ddag$Center for Theoretical Study,
Charles University, Jilsk\'a 1, 110 00 Prague, Czech Republic}
\date{February 1, 2000}

\begin{abstract}
We present a general, rigorous theory of Lee-Yang zeros for models 
with first-order phase transitions that admit convergent contour 
expansions. We derive formulas for the positions and the density 
of the zeros.  In particular, we show that for models without 
symmetry, the curves on which the zeros lie are generically not 
circles, and can have topologically nontrivial features, such as 
bifurcation.  Our results are illustrated in three models in a 
complex field: the low-temperature Ising and Blume-Capel models, 
and the $q$-state Potts model for $q$ large enough. 
\end{abstract}

\maketitle


Almost half a century ago, in two classic papers \cite{YL}, Lee 
and Yang studied the zeros of the Ising partition function in the 
complex magnetic field plane, showed rigorously that the zeros lie 
on the unit circle, and proposed a program to analyze phase 
transitions in terms of these zeros. A decade later, Fisher 
\cite{F} extended the study of the Ising partition function zeros 
to the complex temperature plane. Since that time, there have been 
numerous studies, both exact and numerical, of the Lee-Yang and 
Fisher zeros in a wide variety of models \cite{pool}.  However, 
with a few notable exceptions \cite{rigorpool}, remarkably little 
progress has been made in extending the rigorous Lee-Yang program. 
This is due to the fact that rigorous statistical mechanics has 
relied almost exclusively on probabilistic techniques which fail 
in a complex parameter space.  In this Letter, we adapt complex 
extensions \cite{compPS,fss,BBCKK} of Pirogov-Sinai theory 
\cite{PSZ} to realize the Lee-Yang program in a general class of 
models with first-order phase transitions.

The purpose of this work is threefold.  First, it is of interest 
to establish the mathematical foundation of a program that has 
been so central to statistical physics.  Second, our theory gives 
a novel physical interpretation of the existence and position of 
partition function zeros by relating them to the phase coexistence 
lines in the complex plane. Finally, from a practical viewpoint, 
our theory provides a framework for the interpretation of 
numerical data by allowing explicit, rigorous computation of the 
position of the zeros.  Indeed, we find rigorous results which 
clarify many ambiguities in published data. Specifically, in 
models without an underlying symmetry, we prove that the zeros 
generically do not lie on circles, even in the thermodynamic 
limit. This applies, in particular, to the Blume-Capel and Potts 
models in complex magnetic fields; see \cite{Lee,KC} for heuristic 
studies of these models. We also prove that the curves defined by 
the asymptotic positions of the zeros can have topologically 
nontrivial features, such as bifurcation and coalescence, and show 
that these features correspond to triple (or higher) points in the 
complex phase diagram.

The results to be stated next are rather technical. Roughly 
speaking, they say that, for models with a convergent contour 
expansion, the partition function can be written in the form 
\eqref{1} and its zeros are given as solutions to equations 
\eqref{2a} and \eqref{3a}. Readers unfamiliar with rigorous 
expansion techniques are encouraged to see the concrete examples 
following the main results.

{\it Main Result:} Consider a $d$-dimensional lattice model with 
$n$ equilibrium phases whose interaction depends on a complex 
parameter $z$. Suppose $d\ge2$ and that $z$ is in the region  
(typically, a large disc or the entire ${\mathbb C}$) 
where the model admits a contour representation with strongly suppressed 
contour weights. Under suitable conditions \cite{compPS,fss,PSZ}, 
there are complex functions $f_\ell = f_\ell(z)$, 
$\ell=1,\dots,n$, such that the partition function in a periodic 
volume $V=L^d$ at inverse temperature $\beta$ can be written as 
\begin{equation}
Z^{\text{per}}_L= \sum_{\ell=1}^n q_\ell\, e^{-\beta f_\ell V}
+{\mathcal O}(e^{-L/L_0}e^{-\beta fV}).
\label{1}
\end{equation}
Here $L_0$ is of the order of the correlation length,
$f=\min_k \Re{\mathfrak{e}} f_k$, and $q_\ell$ is the degeneracy of
the phase $\ell$. Physically, 
$f_\ell$ can be interpreted 
as metastable free energies with the stability of the
$\ell^{\text{th}}$ phase being characterized by the condition
$\Re{\mathfrak{e}} f_\ell =f$. If $\ell$ is stable, $f_\ell$ is just the
free energy of the system with boundary condition $\ell$ \cite{complfreen}.
In the region where $\ell$ is not stable, $f_\ell$ is constructed as 
a smooth extension of $f_\ell$ from the stable region. Clearly,
$f_\ell$ depends on the parameters of the model, but not~on~$L$.

Eq.\ \eqref{1} can be used to locate the zeros of 
$Z^{\text{per}}_L$ analytically. Excluding a neighborhood of size 
$\delta_L\!\sim\! L^{-(d-1)}$ of the triple or higher coexistence 
points \cite{triple} and assuming a  degeneracy removing 
condition \cite{nondeg}, each zero of $Z^{\text{per}}_L$ lies 
within ${\mathcal O}(e^{-L/L_0})$ of a solution to the equations 
\begin{equation}
\Re{\mathfrak{e}} f_{\ell,L}^{\eff} = \Re{\mathfrak{e}}
f_{m,L}^{\eff} < \Re{\mathfrak{e}} f_{k,L}^{\eff}  \text{ \ for all
} k\neq \ell,m, \label{2a}
\end{equation}
\begin{equation}
\beta V(\Im{\mathfrak m}f_{\ell}-\Im{\mathfrak m}f_m)
=\pi \text{ mod }2\pi
\label{3a}
\end{equation}
for some $\ell\neq m$, where $f_{\ell,L}^{\eff}=f_\ell-(\beta 
V)^{-1}\log q_\ell$. In fact, the solutions to \eqref{2a} and 
\eqref{3a} and the zeros of $Z^{\text{per}}_L$ are in one-to-one 
correspondence. As a consequence, the zeros of $Z^{\text{per}}_L$ 
asymptotically concentrate on the phase coexistence curves 
$\Re{\mathfrak{e}} f_{\ell}= \Re{\mathfrak{e}} f_m$ with the 
density $\frac1{2\pi}\beta  V |(d/dz)(f_\ell-f_m)|$. Inside the 
$\delta_L$-neighborhood of the multiple coexistence points, both 
the analysis and the resulting equations for the zeros are more 
complicated; see \cite{BBCKK} for details. However, it turns out 
that all but  a uniformly bounded number of zeros (out of the 
total of order $L^d$) can be accounted for by the simple equations 
\eqref{2a} and \eqref{3a}.

The proof, which appears elsewhere \cite{BBCKK}, is technically
complicated, but the main idea is simple.  The key input is a
complex version of methods developed mainly in the context of
finite-size scaling \cite{PSZ,compPS,fss} leading to
equation \eqref{1} and similar expressions for the derivatives
of $Z_L^{\text{per}}$.
Equations \eqref{2a} and
\eqref{3a} for the zeros of $Z_L^{\text{per}}$ arise from
``destructive interference'' of two terms, $q_\ell e^{-\beta
f_\ell V}$ and  $q_m e^{-\beta f_m V}$, in the sum in \eqref{1}.
Outside the $\delta_L$-neighborhood of multiple coexistence
points, all other terms are negligible.

To illustrate our result, we will discuss three specific
models in the presence of a complex external field.

{\it Ising Model:}  The nearest-neighbor Hamiltonian is
$$
\beta H = -J\sum_{\langle x,y\rangle}\sigma_x\sigma_y - h\sum_x \sigma_x.
$$
Here  $\sigma_x\in\{-1,+1\}$,
the coupling $J>0$ is taken large enough to ensure absolute
convergence of the low-temperature expansion, and $h$ is the
complex external field. Neglecting the error term,
Eq.~\eqref{1}
becomes
$$
Z^{\text{per}}_L=e^{-\beta f_+ V}+e^{-\beta f_- V}.
$$
This leads to the following equations for the zeros:
\begin{equation}
\label{2}\Re{\mathfrak e} (f_+-f_-)=0
\end{equation}
\begin{equation}
\label{3}\Im{\mathfrak m} (f_+-f_-)=\textstyle
\frac{(2k-1)\pi}{\beta V},\quad k=1,\dots, V.
\end{equation}
Inserting
the low-temperature expansions of $f_\pm$,
$$
\beta f_\pm=\mp h-dJ-e^{-4dJ}e^{\mp2h}+R(\pm h),
$$
where $R(h)$ and its derivative are both ${\mathcal 
O}(e^{-4(2d-1)J})$, we find that the zeros occur at 
$e^{2h}=e^{i\theta_k}$, with 
$$
\textstyle \theta_k=\frac{(2k-1)\pi}{V} 
+2e^{-4dJ}\sin\Bigl(\frac{(2k-1)\pi}V\Bigr) +{\mathcal O}\bigl(\frac 
kV e^{-4(2d-1)J}\bigr), 
$$
$k=1,\dots, V$. Moreover, the $h\leftrightarrow-h$ sym\-metry can 
be used to prove that condition \eqref{2} is equivalent to 
$\Re{\mathfrak e}\,h=0$, guaranteeing that the zeros of 
$Z^{\text{per}}_L$ lie within an ${\mathcal 
O}(e^{-L/L_0})$-neighborhood of the unit circle. The 
$h\leftrightarrow -h$ symmetry of the partition function 
then allows us to conclude that, for large $L$, the zeros lie {\it exactly\/} on
the unit circle; see the end of the Blume-Capel section for 
details of an analogous argument.
This gives an alternative proof of the Lee-Yang circle 
theorem at low temperatures~\cite{BBCKK}.
We stress that symmetry is the key factor 
here; in the~ab\-sence of symmetry, \eqref{2} does not in general 
lead to circles.

{\it Blume-Capel Model:}
The Hamiltonian~\cite{BC} is
$$
\beta H=-\sum_{\langle x,y\rangle}
J(\sigma_x-\sigma_y)^2-\sum_x (\lambda \sigma_x^2+ h\sigma_x)
$$
with spins $\sigma\in\{-1,0,+1\}$, real parameters $J>0$ 
and~$\lambda$, and complex field $h$. For $J$ large, the real 
$(\lambda,h)$-phase diagram features three phases labeled by $+$, 
$0$, and~$-$, each with an abundance of the corresponding spin. 

The zeros of this model are shown in Fig.~1. Note that the zeros 
have  a non-uniform distribution, forming curves of non-circular 
shape, and that for $\lambda$ in a certain interval 
$(\lambda_c^-,\lambda_c^+)$, bifurcation (i.e., splitting of the 
curve) occurs. In the remainder of this section, we rigorously 
establish these features for large $J$. Before beginning our 
analysis, we remark that in~\cite{Lee} a phenomenological theory 
of partition function zeros based on \cite{fss} was developed and 
then applied  to the Blume-Capel model. In contrast to our 
approach, that of \cite{Lee} gives no quantitative estimate of 
approximations or errors, and it misses certain important 
qualitative features, namely the bifurcation.

\begin{figure}
\centerline{\epsfxsize=3.4truein\epsffile{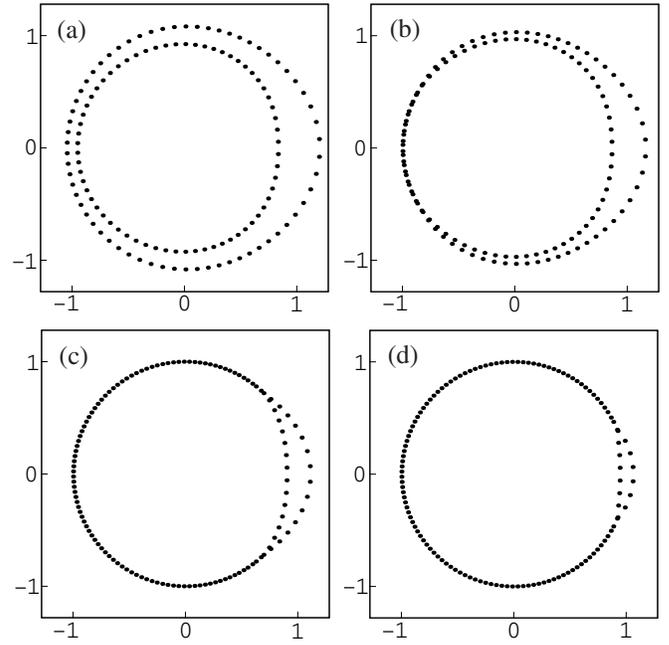}} \vspace{1mm}
\caption{ The 128 zeros of the partition function of the 
Blume-Capel model in the complex $e^{h}$ plane, for the $8\times8$ 
periodic square grid at $e^{-4J}=1/{16}$ and $e^\lambda=$ (a) 
$0.9$, (b) $0.94$, (c) $1$, (d)~$1.07$. The actual zeros lie 
within a distance of order $e^{-10 J}\!\!+\!e^{-L/L_0}$ of those 
depicted. For $\lambda<\lambda_c^-\approx -e^{-4J}$, the outer 
region of the $+$ phase is separated from the inner $-$ phase by 
an annular region of the $0$ phase (a); asymptotically, the zeros 
lie on the boundaries of these regions. As $\lambda$ increases 
through $\lambda_c^-$, the two boundaries coalesce on the left 
hand side, leading to bifurcation for $\lambda>\lambda_c^-$. The 
common boundary  grows (c,d) and, eventually, at 
$\lambda=\lambda_c^+\approx e^{-4J}$, the $0$ phase disappears and 
bifurcation terminates. For $\lambda>\lambda_c^+$, all zeros lie 
on the unit circle. } \vspace{-2mm} 
\end{figure}

Fix $J$ large and let $e^h = z = re^{i\theta}$.  Our analysis is 
done in two steps.  First, we focus on the unit circle, $r=1$, 
and identify $\lambda_c^\pm$ and the position $e^{\pm i 
\theta_c(\lambda)}$ of the splitting points.  Then we extend the 
analysis to all~$r$. 

The $J=\infty$ phase diagram has three ground states, $\sigma 
\equiv  -1$, $0$, $1$, with energy densities $-\lambda+h$, $0$, 
$-\lambda-h$. 
\noindent
The large-$J$~expansions of the free energies are 
$$
\begin{aligned}
e^{-\beta f_+}\!\!&=\!z e^\lambda\exp\bigl\{ {\textstyle\frac1z}
e^{-\lambda-4J}
+2{\textstyle\frac1{z^2}}e^{-2\lambda-6J}
+R_+(z)\bigr\}\\
e^{-\beta f_-}\!\!&=\!{\textstyle\frac1z} e^\lambda\exp\bigl\{ z
e^{-\lambda-4J}
+2z^2e^{-2\lambda-6J}+R_-(z)\bigr\}\\
e^{-\beta f_0}\!&=\!\exp\bigl\{ \!(z\!+\!{\textstyle\frac1z})
e^{\lambda-4J}\!+\!
2(z^2\!+\!{\textstyle\frac1{z^2}})e^{2\lambda-6J}
\!+\!R_0(z)\!\bigr\}\!,
\end{aligned}
$$
which, for brevity, we write only for $d=2$ \cite{BCfreeen}.
Here $R_\pm(z)$ and $R_0(z)$ and their first two derivatives
are all ${\mathcal O}(e^{-8J})$.
There are no phase degeneracies, $q_0=q_\pm=1$.

By Eq.~\eqref{2a}, the analysis of the loci of the zeros requires 
comparison of $\Re{\mathfrak{e}} f_+$, $\Re{\mathfrak{e}} f_-$, 
and  $\Re{\mathfrak{e}} f_0$. On the unit circle, 
$\Re{\mathfrak{e}} f_+ =\Re{\mathfrak{e}} f_-$. Thus it suffices 
to study the sign of $\Delta_\pm=$ $\Re{\mathfrak{e}}\beta 
f_{\pm}-\Re{\mathfrak{e}}\beta f_0$. For $|\lambda| \gg e^{-4J}$, 
$\operatorname{sgn}(\Delta_\pm) =-\operatorname{sgn}\lambda$. So 
let $\lambda = {\mathcal O}(e^{-4J})$. Then, for $J$ large, 
\begin{equation}
\label{6a}
\Delta_\pm=-\lambda+
e^{-4J}(2e^\lambda\!-\!e^{-\lambda})\cos\theta
+{\mathcal O}(e^{-6J}),
\end{equation}
so that $(d/d\theta)\Delta_\pm<0$ for $\theta\in(0,\pi)$ 
\cite{remark} and, similarly, $(d/d\lambda)\Delta_\pm 
=-1+{\mathcal O}(e^{-4J})$. This implies the existence of 
$\lambda_c^\pm = \pm e^{-4J} +{\mathcal O}(e^{-6J})$ and  
$\theta_c(\lambda)$, with $\theta_c(\lambda)\in(0,\pi)$ for all 
$\lambda\in(\lambda_c^-,\lambda_c^+)$, such that, on the unit 
circle, $0$ is the only stable phase for all $\theta$ when 
$\lambda<\lambda_c^-$ and for $|\theta|<\theta_c(\lambda)$ when 
$\lambda\in[\lambda_c^-,\lambda_c^+]$, whereas $\pm$ are the only 
stable phases in the complementary region of $(\lambda, \theta)$. 
Moreover, $\theta_c(\lambda)$ decreases with $\lambda$ and 
$\theta_c(\lambda)\to 0$ (resp.,\ $\pi$) when 
$\lambda\uparrow\lambda_c^+$ (resp.,\ $\lambda\downarrow\lambda_c^-$). 

Now let $r$ be arbitrary. We have $(d/dr) \Re{\mathfrak 
e}f_\pm(z)=\pm r^{-1} +{\mathcal O}(e^{-4J})$ and 
$(d/dr)\Re{\mathfrak e}f_0(z)={\mathcal O}(e^{-4J})$. Using the 
symmetries of the model, 
$$
\Re{\mathfrak e}f_\pm(z)=\Re{\mathfrak 
e}f_\mp(z^{-1})\quad\text{and}\quad \Re{\mathfrak 
e}f_0(z)=\Re{\mathfrak e}f_0(z^{-1}), 
$$
it follows that there is a function $\theta\mapsto r(\theta)$, 
$0\le 1-r(\theta)\le {\mathcal O}(e^{-4J})$, 
$r(\theta)=r(-\theta)$, such that $-$ is stable for $r \leq 
r(\theta)$, $0$ is stable for $r(\theta) \leq r \leq 1/r(\theta)$, 
and $+$ is stable for $r \geq 1/r(\theta)$.  Notice that 
$r(\theta)=1$ for $|\theta|\ge\theta_c(\lambda)$ when 
$\lambda_c^-\le\lambda\le\lambda_c^+$ and for all $\theta$ when 
$\lambda>\lambda_c^+$.

Consider now $L\gg L_0$ and suppose there is a partition function 
zero at $z_0=\rho e^{i\psi }$. If $r(\psi )<1$, then the zero lies 
close to one of the curves defined by the equations 
$\Delta_\pm=0$. We claim that these curves are non-circular and 
that the zeros do not maintain a uniform spacing along them. 
Indeed, set $\lambda=0$ for simplicity and observe that 
$$
e^{\Delta_\pm}
=\bigl|z^{\mp 1}
\exp\bigl\{ z^{\pm 1}e^{-4J}
+2z^{\pm2}e^{-6J}+{\mathcal O}(e^{-8J})\bigr\}\bigr|.
$$
Replacing $z^{\pm1}$ by $x+iy$, the equation $\Delta_\pm=0$ and the
expansion of the exponential up to ${\mathcal O}(e^{-8J})$  yield
$$
x^2+y^2=1+2xe^{-4J}+4(x^2-y^2)e^{-6J}+{\mathcal O}(e^{-8J}).
$$
This is an ellipse centered at $e^{-4J}+{\mathcal O}(e^{-8J})$ 
with semiaxes $1\pm 2e^{-6J} +{\mathcal O}(e^{-8J})$. To determine 
the density of zeros, we compute $|(d/dz)(f_\pm-f_0)|$ and easily 
verify that it is non-constant on the above ellipse. 

If, on the other hand,  $r(\psi )=1$ \cite{triplepoints}, then, 
for $L$ large enough, the zero necessarily lies {\it exactly} on 
the unit circle, $\rho=1$. Indeed, by \eqref{2a}, \eqref{3a}, and 
the degeneracy removing condition \cite{nondeg}, the distance 
between two adjacent zeros is of order $L^{-d}$. But we also have 
$|\rho-r(\psi )| = |\rho - 1| \le {\mathcal O}(e^{-L/L_0})$, and 
if $\rho\not=1$, then by symmetries of the model,  there would be 
another zero at $\bar z_0^{-1}=\rho^{-1}e^{i\psi }$. However, 
$|z_0-\bar z_0^{-1} |\le {\mathcal O}(e^{-L/L_0})\ll L^{-d}$, a 
contradiction. A similar argument  proves a ``local'' version of 
the Lee-Yang theorem \cite{class} in a large class of models for 
which the standard, ``global'' theorem fails.

\paragraph*{Potts Model:\/}
The
Hamiltonian is
\begin{equation}
\label{6}
\beta H=-J\sum_{\langle x,y\rangle} \delta_{\sigma_x,\sigma_y}
-h\sum_x \delta_{\sigma_x,1}.
\end{equation}
with spins $\sigma\in\{1,2,\dots,q\}$, real coupling $J>0$, and 
complex field $h$. For $h=0$ this is the standard Potts model, 
with a $q$-fold degenerate ordered phase at large $J$ and a 
disordered phase at small $J$, coexisting at $J_c^{q}\approx 
\frac1d\log q$. The transition is first-order for large $q$ 
\cite{KS}, while it is presumably second order for $q\le q_c(d)$. 
For $h \neq 0$ and $q$ large, the phase diagram was determined 
first by formal expansion \cite{G}, and recently by rigorous 
probabilistic methods \cite{BBCK}. The Lee-Yang zeros of \eqref{6} 
were studied numerically in \cite{KC}, where it was suggested that 
the zeros lie on almost circular curves slightly outside the unit 
circle,  for $J$ both above and below $J_c^{q}$. While, by 
three-phase coexistence, this turns out to be incorrect for $J< 
J_c^{q}$ (see Fig.~2), we prove that this is indeed the case for 
$J\ge J_c^{q}$, thus resolving a controversy in \cite{KC}. 

\vspace{2mm}
\begin{figure}[t]
\centerline{\hglue-1mm\epsfxsize=3.375truein\epsffile{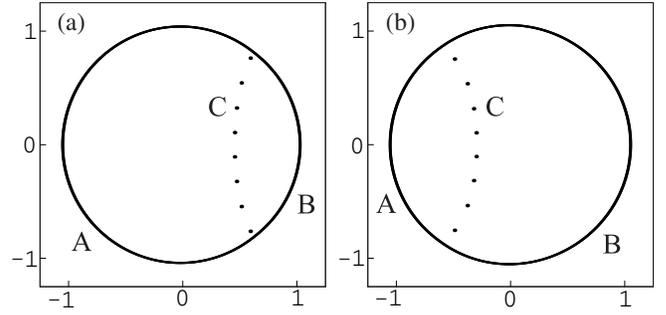}}
\vspace{2mm} \caption{Complex-$e^h$ diagram showing the zeros of 
the  Potts model in a three-dimensional periodic box of size 
$V=1000$ with parameters $q=25$, $e^{3J}/q=$ (a) $1.185$ and (b) 
$1.155$. In each case, there are 1000 zeros distributed on three 
non-circular arcs, labeled A, B, and C, with those on A 
and B denser than those on C. The outer region corresponds to the 
ordered magnetized phase, while the regions left, resp., right, of 
arc C contain the ordered non-magnetized and disordered 
phase. For $V$ large, arc C shows up first at $J=J_c^{q}$, 
it passes through zero at $J=J_c^{q-1}$, and it disappears at 
$J\approx J_c^{q-2}$.} \vspace{-2mm} 
\end{figure}

The model~\eqref{6} has three phases: the disordered phase $(D)$ 
with degeneracy $q_D=1$, and two ordered phases: a magnetized 
$(M)$ and a non-magnetized $(O)$ phase, with  degeneracies $q_M=1$ 
and $q_O=q-1$, characterized by abundances of $\sigma_z=1$ and 
$\sigma_z=\text{const.}\neq1$, respectively. Let us abbreviate 
$z=e^h$, $\kappa_d=d(2d-1)$, $Q_z^{(k)}=q-1+z^k$, and 
$Q_z=Q_z^{(1)}$. The free energies are given \cite{BBCK} by 

\begin{widetext}
\vspace{-6mm}
$$
\begin{aligned}
e^{-\beta f_D}&=
Q_z\exp\bigl\{d(e^J-1)Q_z^{(2)}/Q_z^2
+ \kappa_d(e^J-1)^2Q_z^{(3)}/Q_z^3
- (\kappa_d+1/2)(e^J-1)^2[Q_z^{(2)}/Q_z^2]^2
+ {\mathcal O}(1/q^{3-4/d})\bigr\}
\\
e^{-\beta f_M}&=
z\exp\bigl\{d J
+ e^{-2dJ}(Q_z z^{-1}-1)
+ de^{-(4d-1)J}(Q_z^2+e^J Q_z^{(2)})z^{-2}
- (d+1/2)e^{-4dJ}Q_z^2 z^{-2}
+ {\mathcal O}(1/q^{3-2/d})\bigr\}
\\
e^{-\beta f_O}&=
\exp\bigl\{d J + e^{-2dJ}(Q_z-1)
+ de^{-(4d-1)J}(Q_z^2+e^J Q_z^{(2)})
- (d+1/2)e^{-4dJ}Q_z^2
+ {\mathcal O}(1/q^{3-2/d})\bigr\}.
\end{aligned}
$$
\end{widetext}

The zeros of the periodic Potts partition function are depicted in 
Fig.~2. In particular, for $J_-< J<J_+$ (where $J_+=J_c^q$ and 
$J_-\approx J_c^{q-2}$), the loci do not lie on a single closed 
curve but rather split the complex plane into three pieces, 
corresponding to the regions of stability of the three phases 
above. The number of zeros on the inner arc is roughly $V/(2\pi 
q)$, so one needs to take $V$ quite large and tune $J$ to fall 
inside the narrow window $(J_-,J_+)$ to find any interior zeros. 
This explains why these zeros were not detected in previous 
numerical work \cite{KC}.

Despite their appearance, none of the curves in Fig.~2 is a 
circle. This is verified by finding the coexistence curves 
\eqref{2a} for three distinct pairs $k,\ell\in\{D,M,O\}$. When 
$J\ge J_c^q$, only the phases $M$ and $O$ are relevant, and the 
asymptotic location of the zeros is given by $\Re{\mathfrak e}
f_O=\Re{\mathfrak e}f_M$. For $z=re^{i\theta}$, this easily 
implies 
$$
r=1+q e^{-2d J}(1-\cos\theta)+{\mathcal O}(1/q^2),
$$
so that for $J\ge J_c^q$ and $V\to\infty$, all zeros with 
$|\theta|\gg 1/\sqrt q$ are asymptotically outside the unit 
circle. By invoking arguments similar to \cite{remark}, this 
extends to all $\theta$ \cite{BBCK}. There are two finite-volume 
corrections: an {\it outward\/} shift of order $1/V$ due to 
$f_O>f_{O,L}^{\text{eff}}$ (see Eq.~\eqref{2a}) and an error 
${\mathcal O}(e^{-L/L_0})$ coming from \eqref{1}. Since 
$1/V\gg{\mathcal O}(e^{-L/L_0})$, this proves the initial 
numerical observation in \cite{KC}. 

To make the interesting features clearly visible, Figs.~1 and~2 
were drawn for values of $e^{-4J}$ and $q$ for which we have not 
proved convergence of our expansions. However, as established 
above, all the depicted behaviors indeed occur once $e^{-J}$ (or 
$1/q$) and $e^{-L/L_0}$ are small enough.

In summary, we identify the loci of complex zeros with the complex 
phase coexistence curves.  For particular models, we use this 
identification to map the precise location of these zeros.  We 
find that, in general, the loci are non-uniform and that the 
resulting curves are non-circular; if more that two phases are 
present, the curves also have bifurcation (i.e., splitting) 
points.

The research of R.K.\ was partly supported by the grant GA\v{C}R$\,$201/00/1149.


\begin{references}
\bibitem{complfreen}
Note that, in the complex setting, the free energy generally depends on the
boundary condition.
\bibitem{nondeg} 
Let $e_\ell$
be the energy density of the zero-temperature state giving
rise to the phase $\ell$. Then we require that $|(d/dz)(e_\ell - e_m)|$
be uniformly positive for $\ell \neq m$; see~\cite{BBCKK}.
\bibitem{BCfreeen}
Here, in the first line, $z e^\lambda$ comes from the energy 
density of the $+$ ground state, whereas the first term in the 
exponential accounts for single-spin flips $+\rightarrow 0$ with 
energy cost $4J+\lambda+h$. The second term in the exponential 
comes from simultaneous flips $++\rightarrow00$ of nearest 
neighbors ($2$~accounts for the orientations of the pair). Similar 
reasoning applies to $f_-$. The $0$ ground state admits both 
$0\rightarrow\pm$ flips in its lowest excitations, leading to the 
terms $z+1/z$ and $z^2+1/z^2$ in the third line.
\bibitem{remark}
Indeed, by \eqref{6a}, this is true outside an~${\mathcal 
O}(e^{-4J})$ neighborhood of $0$ and $\pi$. At these points, 
$(d/d\theta)\Delta_\pm=0$ by symmetry. Since 
$(d^2/d\theta^2)\Delta_\pm<0$ (resp.,\ $>0$) in a ${\mathcal 
O}(1)$ neighborhood of $\theta=0$ (resp.,\ $\pi$), 
$(d/d\theta)\Delta_\pm$ is strictly decreasing (increasing) there, 
which proves the claim.
\bibitem{triplepoints}
More precisely, we need that $r(\theta)=1$ for 
$|\theta-\psi|\le\delta_L$.
\bibitem{class}
Our local theorem requires that, in an $\epsilon$-neighborhood 
${\mathcal U}_\epsilon(z_0)$ of some $z_0$ with $|z_0|=1$, (i) 
maximally two phases are stable, (ii) the phases are related by 
the $h\leftrightarrow -h$ symmetry, and (iii) they coexist on a 
curve $z(\theta)=r(\theta)e^{i\theta}$ in ${\mathcal 
U}_\epsilon(z_0)$. Then $r(\theta)\equiv 1$ in ${\mathcal 
U}_\epsilon(z_0)$, and all zeros in  ${\mathcal 
U}_{\epsilon'}(z_0)$ with $\epsilon'=\epsilon-\delta_L$ lie on the 
unit circle. See \cite{BBCKK}.
\end{references}
\end{document}